# Initial results from the New Horizons exploration of 2014 $MU_{69}$, a small Kuiper Belt Object

Corresponding Author: S.A. Stern, astern@boulder.swri.edu

S. A. Stern,[1] H. A. Weaver,[2] J. R. Spencer,[1] C. B. Olkin,[1] G. R. Gladstone,[3] W. M. Grundy,[4] J. M. Moore,[5] D. P. Cruikshank,[5] H. A. Elliott,[3,48] W. B. McKinnon,[6] J. Wm. Parker,[1] A. J. Verbiscer,[7] L. A. Young,[1] D. A. Aguilar,[8] J. M. Albers,[2] T. Andert,[9] J. P. Andrews,[1] F. Bagenal,[10] M. E. Banks,[11] B. A. Bauer,[2] J. A. Bauman,[12] K. E. Bechtold,[2] C. B. Beddingfield,[13,5] N. Behrooz,[2] K. B. Beisser,[2] S. D. Benecchi,[14] E. Bernardoni,[10] R. A. Beyer,[13,5] S. Bhaskaran,[15] C. J. Bierson,[16] R. P. Binzel,[17] E. M. Birath,[1] M. K. Bird,[18,28] D. R. Boone,[15] A. F. Bowman,[2] V. J. Bray,[19] D. T. Britt,[20] L. E. Brown,[2] M. R. Buckley,[2] M. W. Buie,[1] B. J. Buratti,[15] L. M. Burke,[2] S. S. Bushman,[2] B. Carcich,[21,2] A. L. Chaikin,[22] C. L. Chavez,[13,5] A. F. Cheng,[2] E. J. Colwell,[2] S. J. Conard,[2] M. P. Conner,[2] C. A. Conrad,[1] J. C. Cook,[23] S. B. Cooper,[2] O. S. Custodio,[2] C. M. Dalle Ore,[13,5] C. C. Deboy,[2] P. Dharmavaram,[2] R. D. Dhingra,[24] G. F. Dunn,[3] A. M. Earle,[17] A. F. Egan,[1] J. Eisig,[2] M. R. El-Maarry,[25] C. Engelbrecht,[2] B. L. Enke,[1] C. J. Ercol,[2] E. D. Fattig,[3] C. L. Ferrell,[1] T. J. Finley,[1] J. Firer,[2] J. Fischetti,[12] W. M. Folkner,[15] M. N. Fosbury,[2] G. H. Fountain,[2] J. M. Freeze,[2] L. Gabasova,[26] L. S. Glaze,[27] J. L. Green,[27] G. A. Griffith,[2] Y. Guo,[2] M. Hahn,[28] D. W. Hals,[2] D. P. Hamilton,[29] S. A. Hamilton,[2] J. J. Hanley,[3] A. Harch,[21] K. A. Harmon,[15] H. M. Hart,[2] J. Hayes,[2] C. B. Hersman,[2] M. E. Hill,[2] T. A. Hill,[2] J. D. Hofgartner,[15] M. E. Holdridge,[2] M. Horányi,[10] A. Hosadurga,[2] A. D. Howard,[30] C. J. A. Howett,[1] S. E. Jaskulek,[2] D. E. Jennings,[11] J. R. Jensen,[2] M. R. Jones,[2] H. K. Kang,[2] D. J. Katz,[2] D. E. Kaufmann,[1] J. J. Kavelaars,[31] J. T. Keane,[32] G. P. Keleher,[2] M. Kinczyk,[33] M. C. Kochte,[2] P. Kollmann,[2] S. M. Krimigis,[2] G. L. Kruizinga,[15] D. Y. Kusnierkiewicz,[2] M. S. Lahr,[2] T. R. Lauer,[34] G. B. Lawrence,[2] J. E. Lee,[35] E. J. Lessac-Chenen,[12] I. R. Linscott,[36] C. M. Lisse,[2] A. W. Lunsford,[11] D. M. Mages,[15] V. A. Mallder,[2] N. P. Martin,[37] B. H. May,[38] D. J. McComas,[39,3] R. L. McNutt, Jr.[2] D. S. Mehoke,[2] T. S. Mehoke,[2] D. S. Nelson,[12] H. D. Nguyen,[2] J. I. Núñez,[2] A. C. Ocampo,[27] W. M. Owen,[15] G. K. Oxton,[2] A. H. Parker,[1] M. Pätzold,[28] J. Y. Pelgrift,[12] F. J. Pelletier,[12] J. P. Pineau,[40] M. R. Piquette,[10] S. B. Porter,[1] S. Protopapa,[1] E. Quirico,[26] J. A. Redfern,[1] A. L. Regiec,[2] H. J. Reitsema,[41] D. C. Reuter,[11] D. C. Richardson,[29] J. E. Riedel,[15] M. A. Ritterbush,[15] S. J. Robbins,[1] D. J. Rodgers,[2] G. D. Rogers,[2] D. M. Rose,[1] P. E. Rosendall,[2] K. D. Runyon,[2] M. G. Ryschkewitsch,[2] M. M. Saina,[2] M. J. Salinas,[12] P. M. Schenk,[42] J. R. Scherrer,[3] W.

R. Schlei,[2] B. Schmitt,[26] D. J. Schultz,[2] D. C. Schurr,[27] F. Scipioni,[13,5] R. L. Sepan,[2] R. G. Shelton,[2] M. R. Showalter,[13] M. Simon,[2] K. N. Singer,[1] E. W. Stahlheber,[2] D. R. Stanbridge,[12] J. A. Stansberry,[43] A. J. Steffl,[1] D. F. Strobel,[44] M. M. Stothoff,[3] T. Stryk,[45] J. R. Stuart,[15] M. E. Summers,[46] M. B. Tapley,[3] A. Taylor,[12] H. W. Taylor,[2] R. M. Tedford,[1] H. B. Throop,[14] L. S. Turner,[2] O. M. Umurhan,[13,5] J. Van Eck,[2] D. Velez,[15] M. H. Versteeg,[3] M. A. Vincent,[1] R. W. Webbert,[2] S. E. Weidner,[39] G. E. Weigle II,[47] J. R. Wendel,[27] O. L. White,[13, 5] K. E. Whittenburg,[2] B. G. Williams,[12] K. E. Williams,[12] S. P. Williams,[2] H. L. Winters,[2] A. M. Zangari,[1] T. H. Zurbuchen[27]

[1]Southwest Research Institute, Boulder, CO 80302, USA
[2]Johns Hopkins University Applied Physics Laboratory, Laurel, MD 20723, USA
[3]Southwest Research Institute, San Antonio, TX 78238, USA
[4]Lowell Observatory, Flagstaff, AZ 86001, USA
[5]NASA, Ames Research Center, Space Science Division, Moffett Field, CA 94035, USA
[6]Department of Earth and Planetary Sciences and McDonnell Center for the Space Sciences, Washington University in St. Louis, St. Louis, MO 63130, USA
[7]Department of Astronomy, University of Virginia, Charlottesville, VA 22904, USA
[8]Independent Consultant, Carbondale, CO 81623, USA
[9]Universität der Bundeswehr München, Neubiberg 85577, Germany
[10]Laboratory for Atmospheric and Space Physics, University of Colorado, Boulder, CO 80303, USA
[11]NASA Goddard Space Flight Center, Greenbelt, MD 20771, USA
[12]KinetX Aerospace, Tempe, AZ 85284, USA
[13]The Search for Extraterrestrial Intelligence Institute, Mountain View, CA 94043, USA
[14]Planetary Science Institute, Tucson, AZ 85719, USA
[15] Jet Propulsion Laboratory, California Institute of Technology, Pasadena, CA 91109, USA
[16]Earth and Planetary Science Department, University of California, Santa Cruz, CA 95064, USA
[17]Massachusetts Institute of Technology, Cambridge, MA 02139, USA
[18]Argelander-Institut für Astronomie, University of Bonn, Bonn D-53121, Germany
[19]Lunar and Planetary Laboratory, University of Arizona, Tucson, AZ 85721, USA

[20]Department of Physics, University of Central Florida, Orlando, FL 32816, USA
[21]Cornell University, Ithaca, NY 14853, USA
[22]Independent Science Writer, Arlington, VT 05250, USA
[23]Pinhead Institute, Telluride, CO 81435, USA
[24]University of Idaho, Moscow, ID 83844, USA
[25]Department of Earth and Planetary Sciences, Birbeck, University of London, WC1E 7HX, London, UK
[26]University Grenoble Alpes, Centre National de la Recherche Scientifique, Institut de Planétologie et d'Astrophysique de Grenoble, 38000 Grenoble, France
[27]NASA Headquarters, Washington, DC 20546, USA
[28]Rheinisches Institut für Umweltforschung an der Universität zu Köln, Cologne 50931, Germany
[29]Department of Astronomy, University of Maryland, College Park, MD 20742, USA
[30]Department of Environmental Sciences, University of Virginia, Charlottesville, VA 22904, USA
[31]National Research Council of Canada, Victoria, BC V9E 2E7, Canada
[32]Division of Geological and Planetary Sciences, California Institute of Technology, Pasadena, CA 91125, USA
[33]Marine, Earth, and Atmospheric Sciences, North Carolina State University, Raleigh, NC 27695, USA
[34]National Optical Astronomy Observatory, Tucson, AZ 26732, USA
[35]NASA Marshall Space Flight Center, Huntsville, AL 35812, USA
[36]Independent Consultant, Mountain View, CA 94043, USA
[37]Independent Consultant, Crested Butte, CO 81224, USA
[38]Independent Collaborator, Windlesham, England, GU20 6YW, UK
[39]Department of Astrophysical Sciences, Princeton University, Princeton, NJ 08544, USA
[40]Stellar Solutions, Palo Alto, CA 94306, USA
[41]Independent Consultant, Holland, MI 49424, USA
[42]Lunar and Planetary Institute, Houston, TX 77058, USA
[43]Space Telescope Science Institute, Baltimore, MD 21218, USA
[44]Johns Hopkins University, Baltimore, MD 21218, USA
[45]Roane State Community College, Oak Ridge, TN 37830, USA
[46]George Mason University, Fairfax, VA 22030, USA
[47]Independent Consultant, Burden, KS 67019, USA

[48]Department of Physics and Astronomy, University of Texas at San Antonio, San Antonio, TX 78249, USA

## Abstract

The Kuiper Belt is a distant region of the Solar System. On 1 January 2019, the *New Horizons* spacecraft flew close to (486958) 2014 $MU_{69}$, a Cold Classical Kuiper Belt Object, a class of objects that have never been heated by the Sun and are therefore well preserved since their formation. Here we describe initial results from these encounter observations. $MU_{69}$ is a bi-lobed contact binary with a flattened shape, discrete geological units, and noticeable albedo heterogeneity. However, there is little surface color and compositional heterogeneity. No evidence for satellites, ring/dust structures, gas coma, or solar wind interactions was detected. $MU_{69}$'s origin appears consistent with pebble cloud collapse followed by a low velocity merger of its two lobes.

The Kuiper Belt (KB) is a disc shaped ensemble of objects in the outer Solar System beyond the orbit of Neptune, which was discovered in 1992. This is the source region for Jupiter-family comets and contains primordial planetesimals and dwarf planets (e.g., *1*). The 2003 Planetary Decadal Survey ranked exploration of the KB at the top of funding priorities for NASA's planetary program (*2*).

The resultant NASA mission, *New Horizons* (e.g., *3*), flew through and explored the Pluto dwarf planet system in 2015 (*4, 5*). The spacecraft has since continued farther to explore Kuiper Belt Objects (KBOs) and the Kuiper Belt radiation and dust environment (*6*).

The target selected for the subsequent *New Horizons* KBO flyby was (486958) 2014 $MU_{69}$ (hereafter $MU_{69}$, also informally referred to as Ultima Thule). This KBO was discovered in 2014 using the *Hubble Space Telescope* (HST) during a dedicated search for *New Horizons* KBO flyby targets (*7, 8*). Prior to the arrival of *New Horizons*, the only definitive information regarding $MU_{69}$ were its orbit (*8*), its red color (*9*), its ~30 km approximate size (*7*), and that it displayed no detectable variations in its lightcurve (*10*) or large, distant satellites.

$MU_{69}$'s orbit has a semi-major axis $a$=44.6 astronomical units (au), with eccentricity $e$=0.042 and inclination $i$=2.45°, making it a member of the Cold Classical KBO (CCKBO) population (cold in the name here refers to their low dynamical excitation, not surface temperature). CCKBOs are thought to be (i) distant relics formed from the Solar System's original protoplanetary disk and (ii) more or less dynamically undisturbed bodies that, therefore, formed in situ ~4.5 Gyr ago and which have since remained at/close to their current, large heliocentric distances (*11, 12*). Compared to other KB populations, CCKBOs have a redder color distribution (e.g., *13*), and a different size-frequency distribution (i.e., the population of objects as a function of object size) (*14*) and higher average visible albedos than are typical in the KB (e.g., *15*). Additionally, many CCKBOs have satellites (*16*).

Because CCKBOs are dynamically undisturbed from their formation location (or nearly so [*12*]), they have never been significantly warmed above the ambient, radiative equilibrium temperatures of 30 to 60 K in the Kuiper Belt. $MU_{69}$'s small equivalent spherical diameter of ~19 km is insufficient to drive internal evolution long after its formation. Therefore, small CCKBOs like $MU_{69}$ are expected to be primordial planetesimals, preserving information on the physical, chemical, and accretional conditions in the outer solar nebula and the processes of planetesimal formation (e.g., *1, 12*).

*New Horizons* flew closest to $MU_{69}$ at 05:33:22.4 (±0.2 seconds, 1σ) universal time (UT) on 1 January 2019. The closest approach distance of 3538.5±0.2 (1σ) km was targeted to the celestial north of $MU_{69}$'s center; its relative speed past $MU_{69}$ was 14.43 km $s^{-1}$. The asymptotic approach direction of the trajectory was approximately in the ecliptic plane at an angle of 11.6 degrees from the direction to the Sun. The flyby's observation planning details have been summarized elsewhere (*6*). This report of initial flyby results is based on the ~10% of all collected flyby data that had been sent to Earth before 1 March 2019; full data transmission is expected to complete in mid-2020.

*New Horizons* carries a suite of seven scientific instruments (*3*); all were used in the flyby of $MU_{69}$. These instruments are: (i) Ralph, consisting of two components, the Multispectral Visible Imaging Camera (MVIC), a multicolor/panchromatic mapper, and the Linear Etalon Imaging Spectral Array (LEISA), an infrared composition mapping spectrometer; (ii) the Long Range Reconnaissance Imager (LORRI), a long focal length panchromatic visible camera; (iii) the Alice extreme/far ultraviolet mapping spectrograph; (iv) a Radio Experiment (REX) to measure surface brightness temperatures and X-band radar reflectivity; (v) the Solar Wind Around Pluto (SWAP) charged particle solar wind spectrometer; (vi) the PEPSSI (Pluto Energetic Particle Spectrometer Science Investigation) MeV charged particle spectrometer; and (vii) the Venetia Burney Student Dust Counter (VBSDC)—a dust impact detector.

## General properties

$MU_{69}$ (Figures 1A-B) has been found to have a bi-lobate shape with two discrete, unequally sized lobes. Its bi-lobate shape demonstrates that it is a contact binary, i.e., a pair of formerly separate objects now in physical contact with one another. These two lobes are in contact at an annular interface with higher than average surface reflectance, which we refer to as $MU_{69}$'s neck. The larger and smaller lobes of $MU_{69}$ have been informally designated "Ultima" and "Thule," respectively; formal names will be assigned at a later date.

The bi-lobate nature of $MU_{69}$ is reminiscent of known bi-lobate comets that have been imaged by spacecraft (*17-22*). However, $MU_{69}$'s residence in the Cold Classical Kuiper Belt makes clear that its bi-lobate shape must be primordial, making $MU_{69}$ the only unquestionably primordial contact binary explored to date by spacecraft.

$MU_{69}$'s two lobes are discrete, have retained their basic shapes, and do not (at the available resolution) display prominent compression fractures, deformation, or other geological features indicative of an energetic/violent merger. To the contrary, all available evidence indicates that $MU_{69}$ is instead the product of a gentle collision or merger of two independently formed bodies, possibly contacting one another at or even more slowly than their mutual gravitational infall speed, which we estimate to be only several m $s^{-1}$ based on plausible densities (see below).

The observed bi-lobate shape is inconsistent with a recent formation from the collision of two separate, heliocentric CCKBOs because the characteristic relative impact speeds of CCKBOs are currently ~300 m $s^{-1}$ (*23*). Such a collision would have heavily deformed or destroyed the two components of the contact binary. We conclude that the lobes instead most likely formed and merged in a gentle dynamical environment, such as in a localized particle cloud collapse early in Solar System history (e.g. *24, 25*), further indicating that $MU_{69}$ itself is primordial. The similarity in albedos and colors of the two lobes (discussed below) is further evidence for this formation hypothesis.

Images taken during the approach phase show that the rotational period of $MU_{69}$ is 15.92±0.02 hours; this period is consistent with other CCKBO rotational periods *(26, 27, 28*). $MU_{69}$'s rotational pole was found to point at approximately right ascension = 311º, declination = -25º, corresponding to an obliquity of 98º with respect to $MU_{69}$'s heliocentric orbital plane; i.e. $MU_{69}$ currently rotates almost face-on to the Sun. This pole position, and the fact that the spacecraft approach vector was only ~36º from the direction of the KBO's spin axis, resulted in a very low apparent light curve amplitude that was not detected during many weeks of observations from *New Horizons* until shortly before the flyb*y*. Because asymmetric re-radiation (i.e., so-called YORP, or Yarkovsky–O'Keefe–Radzievskii–Paddack) effects are ineffective on bodies the size of $MU_{69}$ at large distances from the Sun (*29, 30*), and because (as described below) the visible evidence for impact cratering on $MU_{69}$ is modest, the spin period and obliquity of $MU_{69}$ are unlikely to have substantially changed since the binary merger.

We simultaneously fitted the shape of $MU_{69}$ and its pole position using a forward-modeling process. For each of the early-return LORRI and MVIC images in which $MU_{69}$ was resolved, this process rendered the shape with the relevant illumination and orientation using the photometric parameters described below, convolved that rendered image with the appropriate point-spread function (for LORRI or MVIC, as appropriate) to simulate the angular resolution and smearing of the real images, and then numerically compared the result to all of the resolved images of $MU_{69}$ that had been returned. The shape model was parametrically defined, separately for each lobe, using the "octantoid" formalism (*31, 32*; see Methods). Our best fitting shape model (Figures 2A-E) has overall dimensions of approximately 35×20×10 km, with estimated uncertainties likely to be less than 1×1×3 km. The thickness (i.e., the third dimension) is the least constrained due to the spacecraft approach being at a high negative (i.e., southern) latitude on the object, meaning that little of the positive (i.e., northern) half of $MU_{69}$ was directly observed.

From this shape model, the Ultima lobe was found to be lenticular with dimensions of approximately 22×20×7 km (uncertainty $\lesssim$0.6×1×2 km), whilst the Thule lobe is more equidimensional, approximately 14×14×10 km (uncertainty $\lesssim$0.4×0.7×3 km). The fitted volumes of Ultima and Thule are 1400 km$^3$ (±<600 km$^3$) and 1050 km$^3$ (±<400 km$^3$), respectively. The centers of the lobes are 16 km apart. The principal axes of the lobes are roughly parallel to one another (likely <7° misalignment), strongly suggesting the lobes tidally locked prior to contact (see below).

The apparently lenticular shape of the Ultima lobe is unlike any other known heliocentrically orbiting Solar System body, but is reminiscent of some small ring-satellites of Saturn, such as Atlas and Pan, which have accreted equatorial ridges of fine-grained material (*33*). The origin of this flattened shape of Ultima is indeterminate, but possible explanations include accretion from a thin (i.e., flattened) particle sheet during pebble cloud collapse (*24*), to the head-on collision of two very similarly sized bodies within a narrow range of speeds (*34*), or due to deformation arising from rapid rotation or tidal forces prior to its merger with Thule. These possibilities are discussed further below.

Because no satellites of $MU_{69}$ were detected (also see below), $MU_{69}$'s density is poorly constrained. Cometary nuclei are the Solar System bodies likely to be most similar to $MU_{69}$; they are generally found to have low density (<$10^3$ kg $m^{-3}$) and highly (>50%) porous (*21*). The most accurately determined comet nucleus density is that of 67P/Churyumov-Gerasimenko, at 533±6 kg $m^{-3}$ (*35*), though how much that comet's density has evolved from its primordial value is unknown. Adopting the shape model in Figure 2, a lower bound on $MU_{69}$'s density of ≈280 kg $m^{-3}$ can be derived assuming $MU_{69}$'s two lobes are only marginally bound by self-gravity and there is no tensile strength in the neck. Alternatively, assuming a characteristic cometary density of 500 kg $m^{-3}$ (*21*) gives a breakup rotation rate for $MU_{69}$ of ≈12 hours. At this density, if the two lobes merged from a mutual circular orbit, substantial loss of angular momentum would have been required to reach their current, 15.92-hour period. Because mutual gravity would exceed centrifugal acceleration, densities greater than ≈280 kg $m^{-3}$ also imply that the neck region is currently under compression.

We calculated the gravitational and rotational potential across the surface of $MU_{69}$ assuming the shape model in Figure 2, a uniform density of 500 kg $m^{-3}$, and its 15.92-hour rotation period. The surface acceleration ranges between 0.5 and 1 mm $s^{-2}$ (and is nowhere negative). The local acceleration slopes (i.e., the gradients of the gravitational and rotational potential across the surface) are low except in the neck where they can exceed 35°. The equators of Ultima and Thule are gravitational and rotational potential highs.

## Surface reflectivity, color, and composition

Figure 3A shows a visible wavelength I/F (reflected intensity divided by incident flux) contour map from the close approach 04 (CA04) LORRI observation (13° solar phase angle) and its corresponding I/F histogram. At this solar phase angle, the modal I/F of $MU_{69}$ is 0.078. The I/F distributions of $MU_{69}$'s two lobes have the same mode within the measurement error of ±0.05 I/F. I/F variations ranging from ~0.02 to ~0.12 are seen on both lobes (including their terminator regions). However, away from the terminator (i.e., the surface boundary between daylight and nighttime where lighting effects bias results), the minimum I/F is ≈0.05. This occurs on a portion of the wall of the large depression on Thule, which we informally refer to as Maryland (surface features on Ultima Thule are informally named after

U.S. states with major contributions to *New Horizons*); see again Figure 1. The maximum I/F of ≈0.12 occurs at both of the bright spots in Maryland, and on the neck region. The I/F distribution of the Ultima lobe has a sharper peak than that of Thule (see Figure 3B).

The brighter material on Ultima and Thule is mostly segregated into three kinds of surface manifestations: (i) nearly circular spots that become more numerous with decreasing size (ranging from a few km across down to the resolution limit of several tens of meters); (ii) curvilinear and quasi-linear features that are narrow compared to their length; and (iii) broad patches, which are more prominent on the Thule lobe. Darker than average surface units on the binary also come in types (i) and (iii), but not (ii).

The origin of these three types of comparatively bright terrain and how they may differ is not clear. However, initial stereographic analysis of *New Horizons* images using established techniques (*36, 37*) shows that many of the observed bright areas on $MU_{69}$ are located either within topographic depressions (e.g., the neck, the bright spots in Maryland, and the floors of pits and troughs), or at the bases/inflection points of slopes. Therefore a possible explanation is that bright, fine-grained material has been transported downslope to these locations, in which case the higher brightness could be due to predominately smaller particle size (*38*). However, other interpretations including compositional or thermal effects, space weathering, and cold trapping are also possible.

We also point out that the origin of the neck feature at the Ultima-Thule merger zone might not have the same origin as other bright locales. Possibilities include (i) surface processes such as fine particle accumulation as discussed above; but also (ii) processes related to the lobe merger, such as the extrusion of preexisting bright surface material on one lobe during the impact; (iii) post-impact thermal extrusion of ices due to changes in thermal properties or conditions at the merger zone after contact; and (iv) evolutionary processes such as preferential space weathering effects or thermal effects created by the geometry surrounding the neck.

Turning to another topic, analysis of the approach and departure images have allowed us to characterize the solar phase curve of $MU_{69}$ at phase angles up to 153°, considerably (~70x) higher than possible for any Earth-based observations (Figure 4). The derived phase coefficient, or slope, is $\beta$=0.038±0.014 magnitudes per degree between 1.3° and 32.5°. This slope is consistent with that measured for other low albedo, small Solar System bodies including comet nuclei (e.g., *39*). Applying a Hapke photometric model (*38*) to the entire phase curve yields the nominal photometric properties of $MU_{69}$'s surface: single scattering albedo $\omega_o$=0.24, mean topographic slope angle $\theta$=33°, and single particle phase function parameters using a McGuire-Hapke formalism (*39*) of $b$=0.32 and $c$=0.75. These results give $MU_{69}$'s visible (0.55 µm V band) geometric albedo (i.e., the albedo at 0° solar phase) of $p_V$=0.165±0.01. This is a typical value for CCKBOs, the geometric albedos of which range from 0.09 to 0.23, with a mean of 0.15 (*40*). $MU_{69}$'s phase integral is $q$=0.37±0.16, yielding a spherical (also called Bond) albedo of 0.061±0.026.

MVIC color images reveal a globally averaged visible wavelength reddish reflectance slope of 31.1±0.5% per 100 nm, computed using MVIC's Blue, Red, and Near-IR filters (Figures 5A-C), where the quoted uncertainty here is statistical only. This color is consistent with that of other CCKBOs (*41, 42, 43*). Only subtle color (and spectral, see below) differences are detected between the two lobes of $MU_{69}$, despite their distinct shapes, appearances, and the two lobes are clearly resolved from one another at the resolution of the color data. Observations of KBOs with satellites show near-equal colors of the orbiting bodies, interpreted as resulting from co-accretion from a locally homogeneous portion of the nebula (*44*).

The most notable regional color and spectral signatures across $MU_{69}$'s surface appear (i) at the neck between the two lobes, and (ii) at the Thule lobe bright spots in Maryland, which display spectral slopes of 28.2±0.2% per 100 nm, and 30.8±0.2% per 100 nm, respectively. At least two locations on the Ultima lobe also show less red than its average color. Principal component analysis shows that 97% of the variance in MVIC color data is attributable to shading and albedo, while image noise and true color contrasts account for just 3%. The subtle color differences seen could be indicative of compositional differences, although differences in particle size, porosity, etc., can also produce differences in spectral slopes.

LEISA spectral observations (see Figure 5D; 45; see Methods) show that $MU_{69}$ is brighter in the near infrared than it is in the visible, demonstrating that the red slope observed using MVIC extends into the IR. From 1.2 to 2.5 µm, the observed I/F after radiometric calibration ranges from 0.15 to 0.2. The colorant responsible for the redness of $MU_{69}$ and other CCKBOs could be tholin-like complex organic macromolecules, produced from simpler species through radiolytic and photolytic breaking of bonds leading to recombination into progressively heavier molecules (e.g., *46, 47*). Space weathering of silicates can also produce a red coloration, but no direct evidence of silicates is seen on $MU_{69}$. Principal component analysis shows that over 90% of the variance in the LEISA data is also due to shading and albedo, with relatively little variance attributable to regional spectral variability.

As shown in Figure 5D, there are key color slope similarities between the spectrum of $MU_{69}$ and those of the KBO (55638) 2002 $VE_{95}$ (*48*) and the escaped KBO 5145 Pholus (e.g., *49*). Also these objects all exhibit an absorption band near 2.3 µm, tentatively attributed to methanol ($CH_3OH$) or perhaps more complex organic molecules intermediate in mass between simple molecular ices and tholins (*50*). Similar spectral features are also apparent, we note, on the large, dark red, equatorial region of Pluto informally called Cthulhu (e.g., *51*), suggesting similarities in the chemical feedstock and processes that could be at work there and on $MU_{69}$.

Broad spectral absorption features on $MU_{69}$ near 1.5 and 2.0 µm indicate the presence of $H_2O$ ice. However, the shallowness of these features suggests that water ice may have a relatively low abundance in $MU_{69}$'s uppermost surface, at least compared with water-ice rich planetary satellites, and even on Pholus and in Cthulhu.  In this regard, opaque species such as complex organics are known for their ability to mask the spectral signature of $H_2O$ ice in the near infrared (*52*). No unambiguous spectral signatures of silicates, or volatile ices like CO, $N_2$, $NH_3$, $CO_2$, or $CH_4$ that were observed on Pluto (*4*) have been detected for $MU_{69}$, but most of these are supervolatile species are not expected at $MU_{69}$ owing to the ready thermally driven escape of such ices over time from this object. An apparent absorption at 1.8 µm has not yet been assigned an molecular identification.

## Thermal considerations

Temperatures on $MU_{69}$ are set by the balance between the absorption of sunlight and thermal emission back to space. For the 6% Bond albedo estimated above, and an assumed 90% emissivity, absorption and emission at $MU_{69}$'s mean distance from the Sun are balanced at 42 K. For plausible values of thermal conductivity in the $10^{-5}$ J $m^{-1}$ $s^{-1}$ $K^{-1}$ range, we estimate the characteristic skin depths to which diurnal and seasonal waves propagate to be ~0.001 m and ~1 m, respectively (53; see Methods). So diurnal and seasonal temperature variations likely only affect the outer few mm to meters of $MU_{69}$'s surface, so the average temperature pertains to the vast majority of $MU_{69}$'s interior. At the 42 K temperature, frozen volatile species such as CO, $N_2$, and $CH_4$ not trapped in clathrates would sublimate and escape relatively rapidly compared to the age of the Solar System, but amorphous $H_2O$ ice would not crystallize, and so could survive over the age of the Solar System.

Near the surface of $MU_{69}$, temperature varies on seasonal and diurnal timescales. Insolation only differs by 17% over the course of $MU_{69}$'s orbit due to its low orbital eccentricity. However, summer/winter hemisphere insolation varies considerably over $MU_{69}$'s 293 year orbit due to the 98° obliquity of its pole, resulting in long polar days and nights during which regions receive continuous sunlight, or none at all, for many decades. Around equinox, the 15.92-hour rotation period would also produce strong diurnal variations in insolation. Changing insolation generates thermal waves as heat is conducted into and out of the subsurface.

We expect summer surface temperatures to approach a maximum instantaneous equilibrium temperature of about 60 K, while winter temperatures (as on the unilluminated face of $MU_{69}$ during the *New Horizons* flyby) are much lower, down to the seasonal skin depth. A preliminary analysis of the REX radiometry data at the 4.2 cm X-band radio wavelength indicates a brightness temperature in the 20 to 35 K range on the winter (night) side of $MU_{69}$. Thermal emission at cm wavelengths emerges from a range of depths, potentially up to many tens of wavelengths below the surface, so the warmer subsurface is an important source of flux. For typical icy satellite emissivity at cm wavelengths (*54, 55*), the observed REX brightness temperature range appears to be consistent with the expected low thermal inertia values estimated for other KBOs (*56*).

## Geophysical and geological properties

We measured the limb profile of $MU_{69}$ using previously published techniques (*57*). These limb profiles (Figure 6A and 6B) were derived from the CA04 LORRI (characteristic 140 m per pixel) and CA06 MVIC (characteristic 130 m per pixel) observations, and compared with best-fitting ellipses for the projected shape of each lobe. The elliptical axes remove the long wavelength signal and are not necessarily representative of the 3-D body shape (i.e., they are not measured with respect to the shape model in Figure 2).

On the larger lobe Ultima, the limb topography shows total (i.e., minimum to maximum deviation) relief of ~1 km, while the relief on the smaller lobe Thule is more muted at ~0.5 km. Assuming a density $\rho$=500 kg m$^{-3}$, a surface gravity $g$=0.001 m s$^{-2}$, and $h$=0.5–1 km topography of vertical scale $h$ implies stresses of order $\rho g h/3$, or ≈100 Pa. Such modest stresses can be supported frictionally. Excessively steep regions (>35°) near $MU_{69}$'s neck likely require additional internal strength to remain stable. Cohesion of several hundreds of Pa would be capable of supporting these slopes. Such strengths are thought to be normal for comet-like bodies (*22*) and so make general sense for KBOs at $MU_{69}$'s size as well.

A geomorphological map of $MU_{69}$ is also shown in Figure 6C. The two lobes have somewhat different surface geology.

Thule's surface is dominated by an ~7 km diameter depression of probable impact origin (unit labeled *lc*, see Fig. 6C for the meaning of these abbreviations), mentioned above. Stereographic analysis shows the depth of Maryland is <2 km; this depth is consistent with the observed limb topography variations. Maryland's interior shows no unambiguous signs of horizontal layering, but does contain two prominent bright spots of similar size and albedo. Two distinct, km-scale, possible impact craters occur on Maryland's rim-crest. Separately, four distinct troughs appear near the terminator of Thule in unit *um*.

Apart from Maryland, the rest of Thule's observed surface is characterized by broad (few km wide), dark swaths (unit *dm*) that separate lighter-toned, mottled units (units *pm*, *um*, and *rm*). In some places, these dark swaths contain bright spots and a bright-floored, quasi-linear trough. The crenulated boundaries of unit *dm* may be a decrescence morphology, whereby this unit is partly bounded by scarps that have retreated. Unit *dm* may be a deposit of volatile ice with bounding scarps forming due to sublimation at its periphery, with the upper surface of the deposit being protected from sublimation by a dark, refractory mantle, perhaps derived from the deposit itself (*58, 59*). Portions of unit *um* that are proximal to Maryland may be ejecta from the crater, or related to ejecta, but this cannot be confirmed with the current analysis. A distinct, relatively bright region (unit *rm*) at the equatorial, distal end of Thule exhibits roughness at the scale of a few hundred meters; some of the features there appear to be pits, craters, or mounds.

A lightly cratered surface of $MU_{69}$ was predicted by a recent cratering model (*23*); and indeed few definitively impact-related scars are identified on $MU_{69}$. We have considered several hypotheses for the origin of the many pits seen near the terminator in Fig. 2A**.** These include structurally controlled collapse pits, outgassing pits, sublimation pits, and impact craters (*60*); they likely are not all created by the same process. Our assessment is that the chains of similarly sized pits are more likely to be formed by internal processes than cratering, but the isolated pits that show approximately circular planform outlines, bowl-shaped interior depressions, and, in some cases, raised rims, are more consistent with impact crater morphology. Quite surprisingly, there are no obvious crater candidates that are intermediate in size between these ~1 km diameter pits (unit *sp*) and Maryland (unit *lc*, ~7 km.

On Ultima, eight similarly sized (~5 km scale) units of rolling topography (based on subtle albedo gradations and limb profiles) dominate its observed landscape (units *ma* to *mh*). The albedo and texture of these units are generally similar to one another, though each contains brighter material to differing degrees. These units abut each other, typically bordered with distinct, curvilinear, generally higher reflectance boundary regions (particularly unit *mh*, which is ringed by a brighter annulus). In one instance (near the terminator), a trough and a short pit chain mark such a boundary, but due to the low solar incidence angles prevailing across much of Ultima, it is unclear if the boundaries are always associated with topographic features. Stereographic analysis indicates that most of these units have broadly positive relief, although the central unit *mh* is relatively flatter. It is unclear if these boundaries necessarily imply superposition relationships amongst the components. One unit, at the limb of Ultima (unit *md*), appears to be more angular in stereography and is either more elevated or tilted with respect to the other components.

The apparent similarity in the size of Ultima's units (*ma* through *mh*) is likely a clue to their origin. Whether they are a relic of Ultima's formation, or a result of a later evolution, is unclear. One formation-related hypothesis is that the individual units on Ultima are accretional subunits of smaller planetesimals that formed Ultima. This apparent assemblage of units on Ultima consistent with observation of and proposals for the formation of layers on comets such as 9P/Tempel and 67P/Churyumov-Gerasimenko (*20, 61*). However, challenges to this hypothesis are the apparently (i.e., at the available resolution) nearly unimodal size of these units and their apparent absence from Thule, though the latter could be the result of resurfacing caused by the Maryland cratering event. Also, at the expected impact speeds during accretion, such as in a local particle cloud collapse resulting in a body like $MU_{69}$ (perhaps no more than a few m $s^{-1}$, based on the mutual escape speed of the lobes), these accretional subunits would likely not be expected to merge into as compact a body as Ultima unless they were extremely weak (i.e., cohesionless and frictionless) at the time of their accretion (*22*).

## Satellites and orbiting rings/dust search

Both satellites and rings have been detected around KBOs larger than $MU_{69}$ and around Centaurs (*16, 62, 63*), that have escaped the KB and now orbit among the giant planets. Although no KBO as small as $MU_{69}$ is known to have rings, satellites around KBOs are common, particularly in the CCKBO population (*16*).

We searched for satellites and rings of $MU_{69}$ using coadded stacks of 10 to 30 second exposure LORRI images acquired during approach to $MU_{69}$. High-resolution images taken near closest approach provide additional constraints on satellites close to $MU_{69}$. No satellites were detected in our data. Figure 7A shows the quantitative limits on satellites obtained from these data as a function of distance from $MU_{69}$.

Additionally, small particles orbiting $MU_{69}$ could form ring or other dust structures with unusual geometries (*64*) because $MU_{69}$'s very weak gravity is of similar magnitude to the solar radiation pressure. However, larger grains, which are less affected by solar radiation pressure, could form a conventional equatorial ring (*64*).

Approach images constrain any ring or dust assemblages located $\gtrsim$250 km from $MU_{69}$ to have I/F $\lesssim 2\times10^{-7}$ (for a 10 km wide ring) at a phase angle of 11° (Figure 7B). This is equal to or less than the I/F's of many faint outer rings of the giant planets which have ring widths >20 km (*65, 66*). A single high phase (165°), forward-scattered observation downlink to Earth so far also shows no evidence for rings or dust farther than ~400 km from $MU_{69}$ at I/F's >$10^{-4}$. The *New Horizons* VBSDC dust counter experiment reported zero dust impact detections during the gravitational stability sphere of $MU_{69}$, consistent with a lack of extant rings or other dust assemblages.

## Exosphere and heliospheric interaction searches

Due to $MU_{69}$'s small size, it is likely that highly volatile ices that might once have been present at its surface would have escaped to space long ago (*6, 60*). However, less volatile ices (e.g., methanol, acetylene, ethane, and hydrogen cyanide) could be retained over geological time scales and irradiation of these species could result in reddening over time as longer chain tholins are produced (*67, 68*). This implies a slow loss of hydrogen atoms to space as surface ices are converted into tholins. In addition to this escaping H flux, occasional large impacts could provide a source for a transient atmosphere.

We searched for evidence of both coma of escaping gas and charged particle emissions from $MU_{69}$. The searches included using the Alice ultraviolet spectrometer to search for resonance line emission from a coma, and in situ searches for emitted $MU_{69}$ ions with SWAP and PEPSSI.

An Alice count rate spectrum is shown in Figure 8A. This 300 s observation was made ~90 minutes before closest approach, from a range *r*~80,000 km. We fitted a model to the spectrum including background emissions from interplanetary hydrogen plus four nearby stars. No coma emissions from $MU_{69}$ were detected. At the brightest likely coma emission of the H 121.6 nm line, we find a 3σ upper limit source rate of $<3\times10^{24}$ H atoms $s^{-1}$ released by $MU_{69}$. Scattering of solar hydrogen light by H atoms at this source rate would produce a detectable emission, assuming a fall off of $r^{-2}$ from $MU_{69}$.

No structured magnetic interaction between the solar wind and $MU_{69}$ was expected because, at a typical interplanetary magnetic field of 0.2 nT, the gyro-radius of a proton picked up by the solar wind is ~$2\times10^4$ km; ~100× larger than $MU_{69}$ and also larger than the flyby closest approach distance of ~3500 km. Figures 8B-E show SWAP data taken during the time of the flyby. The bulk solar wind density and speed measured by SWAP near $MU_{69}$ are ~2000 protons $m^{-3}$ and ~425 km $s^{-1}$, respectively, for a solar wind flux of $8.5\times10^8$ protons $m^{-2}$ $s^{-1}$. Variations are present due to changes in spacecraft orientation relative to the solar wind, but there is no signature of any detected interaction of the solar wind with $MU_{69}$. Figures 8F-I show PEPSSI data taken during the flyby, which also show no evidence of any $MU_{69}$-related signature. All PEPSSI variations in count rate are, as with SWAP, associated with spacecraft attitude changes, and are consistent with an unperturbed interplanetary medium.

We can estimate the interaction that might be detected by SWAP or PEPSSI using the upper limit from Alice. For an outflow source rate of $Q$ particles $s^{-1}$, the density at $r$ is given by $n \sim Q/4\pi r^2 v$. For the Alice upper limit of $Q \sim 3\times10^{24}$ H atoms $s^{-1}$, the density at closest approach range is $\sim 2.4\times10^{7}$ H atoms $m^{-3}$ (assuming the H atoms have a radial velocity $v \sim 800$ m $s^{-1}$, their thermal speed at 40 K). A fraction $\gamma$ of those H atoms can become ionized and picked up by the solar wind to be detected by SWAP or PEPSSI. Then the count rate is $R = \varepsilon G n v_{SW} \gamma / 4\pi$, where $\varepsilon$ is the detection efficiency near the solar wind speed, $G$ is the instrument geometric factor, and $v_{SW}$ is the solar wind speed. This indicates a PEPSSI count rate of $R \sim 8000\ \gamma$ counts $s^{-1}$. We estimate $\gamma \sim 2\times10^{-7}$ from the fraction of H atoms that could be photo-ionized by sunlight during the travel time $t \sim d_{CA}/v$ for H atoms to get from $MU_{69}$ to the closest approach distance of New Horizons ($d_{CA}$); the expected PEPSSI count rate of $1.6\times10^{-3}$ Hz would thus be ~600 times smaller than the typical background count rate (~1 Hz). A similar result is found for SWAP. Hence the upper limit found with Alice data is more constraining than the SWAP and PEPSSI upper limits.

For comparison, we estimate the loss rate from photosputtering of water ice on $MU_{69}$ to be $\sim 10^{19}$ H atoms $s^{-1}$, much less than our detection upper limit. This estimate is based on the combined direct solar and interplanetary medium H 121.6 nm flux at $MU_{69}$ of $1.6\times10^{12}$ photons $m^{-2}$ $s^{-1}$ and $4.0\times10^{11}$ photons $m^{-2}$ $s^{-1}$, respectively, (*69, 70*) and a yield of 0.003 for H 121.6 nm sputtering of water ice (*71, 72, 73*). Projected and total surface areas for $MU_{69}$ of $\sim 4.1\times10^{8}$ $m^2$ and $1.7\times10^{9}$ $m^2$, respectively, were used in this estimate. Over 4.5 billion years the water ice lost by this process would reduce the size of $MU_{69}$ by a negligible ~0.01 m. Water ice is also eroded by solar wind ions (mainly protons), but this process is estimated to remove only about the same amount of material over time as the photo-sputtering (*74*).

## Implications for formation

$MU_{69}$ has revealed many fascinating aspects and also some puzzles. The latter include the origin of its two non-spherical and markedly different lobe shapes, the provenance of its brighter spots, zones, and linear/curvilinear features, the nature of the similarly-sized surface units on Ultima, the degree to which the object is cratered, the origin of its bright neck, and how its two lobes formed and then merged to create the contact binary we observe.

Despite these puzzles, $MU_{69}$ has already provided information on the ancient accretion processes that operated in the distant protosolar nebula and the Kuiper Belt (*75, 76*). For example, $MU_{69}$ lends weight to model predictions that binaries in the Kuiper Belt may have formed in local, low- to medium-velocity accretion clouds as in pebble cloud gravitational collapse models (*24*). The lack of strong surface albedo, color, and composition heterogeneity between the two lobes supports this hypothesis, because in a local pebble cloud collapse, $MU_{69}$'s two lobes would form from a single source of material. This being said, it is also possible that material accreted throughout the CCKB may have been compositionally homogeneous to begin with.

The binary size ratios produced in existing pebble cloud collapse models (*24*), while focused on the formation of much more massive (100-km scale), co-orbiting binaries, are also consistent with the size ratio of the two lobes of $MU_{69}$ (size ratio ≈0.75). Such models also produce appropriately low merger speeds (*24, 25*) if scaled virially to much less massive particle clouds. However, $MU_{69}$'s pole is highly inclined to the ecliptic, which is not a common outcome in cloud collapse models, given that the cloud's initial rotation state is set by heliocentric Keplerian shear. In contrast, turbulent particle concentration models, such as an over-dense collection of swirling particles collapsing under the influence of aerodynamic drag in the protosolar nebula (*77*), have no preferred initial swirl (mean angular momentum) orientation.

Fundamentally different mechanisms to local pebble or particle cloud collapse have also been proposed for the formation of small-body binaries. However, some such mechanisms, such as YORP spin-up and fission (*29, 30*), apply only in the inner Solar System where thermal radiation forces are sufficiently strong. In the outer Solar System, binary systems may instead form via three-body exchange capture (*78*), but such mechanisms require heliocentric encounter velocities generally near or lower than the Hill speed (the heliocentric Keplerian shear speed at the limit of the primary body's gravitational sphere of influence [*24*]), which for $MU_{69}$ would have been an implausibly low ≈1 cm $s^{-1}$; hence such models are disfavored. We also do not favor this three-body formation mechanism because binaries formed by such three-body exchange would likely rotate either prograde or retrograde (*24, 79*), not highly obliquely like $MU_{69}$.

For $MU_{69}$'s two lobes to reach their current, merged spin state, they must have lost angular momentum if they initially formed as co-orbiting bodies. The lack of detected satellites of $MU_{69}$ may imply ancient angular momentum sink(s) via (i) the ejection of formerly co-orbiting smaller bodies by Ultima and Thule, or (ii) by gas drag, or both. This suggests that contact binaries may be rare in CCKBO systems with orbiting satellites. Another possibility however is that the lobes Ultima and Thule impacted one another multiple times—shedding mass along with angular momentum before making final contact. But the alignment of the principal axes of $MU_{69}$'s two lobes tends to disfavor this hypothesis. In contrast, tidal locking could quite plausibly have produced the principal axis alignment we observe, once the co-orbiting bodies were close enough and spin-orbit coupling was most effective (*80*). Gas drag could also have played a role fostering the observed co-planar alignment of the Ultima and Thule lobes (Figure 2). Post-merger impacts may have also somewhat affected the observed, final angular momentum state.

## Supplementary Materials

Materials and Methods

References (*84-87*)

## Acknowledgements

**Acknowledgements**: We thank the entire, ~2500 person, present and past *New Horizons* team, NASA, and NASA's the Deep Space Network, KinetX Aerospace Corporation, the Caltech Jet Propulsion Laboratory, and the European Space Agency *Gaia* and NASA *HST* space missions for their contributions to making the flyby of $MU_{69}$ successful. We acknowledge the contributions of and thank our NASA Headquarters Program Scientist, Curt Niebur, Jay Anderson of the Space Telescope Science Institute, and of our late team members Thomas Coughlin, Bob Farquhar, Bill Gibson, Lisa Hardaway, and David C. Slater. We thank NASA Administrator Jim Bridenstine for his key support during the December 2018 to January 2019 partial US government shutdown. We also thank SwRI President Adam Hamilton and Johns Hopkins Applied Physics Laboratory Director Ralph Semmel for many years of valuable project support. Finally, we thank the three anonymous referees and editor Keith Smith for their helpful contributions to this paper. **Funding**: This work was supported by NASA's *New Horizons* project via contracts NASW-02008 and NAS5-97271/TaskOrder30. JJK acknowledges funding provided by the National Research Council of Canada. **Author contributions:** SAS, JRS, JMM, OLW, WBM, WMG, GRG, HAE were responsible for drafting this manuscript. SAS is the Principal Investigator of the *New Horizons* mission. All other authors participated either in mission planning, mission operations, mission engineering, mission management, mission public affairs, or science data reduction or analysis, or provided inputs/critique to this manuscript. Coauthor Heather Elliot acknowledges that she was on the NASA Heliophysics Advisory Committee. **Competing interests**: We declare no competing interests. **Data and materials availability**: All images, spacecraft data, and the shape model used in this paper are available at http://dx.doi.org/10.6084/m9.figshare.7940630 . Additional fully calibrated New Horizons $MU_{69}$ data and higher order data products will be released by the NASA Planetary Data System at https://pds-smallbodies.astro.umd.edu/data_sb/missions/newhorizons/index.shtml in a series of stages in 2020 and 2021, owing to the time required to fully downlink and calibrate the dataset.

# Figures and Captions

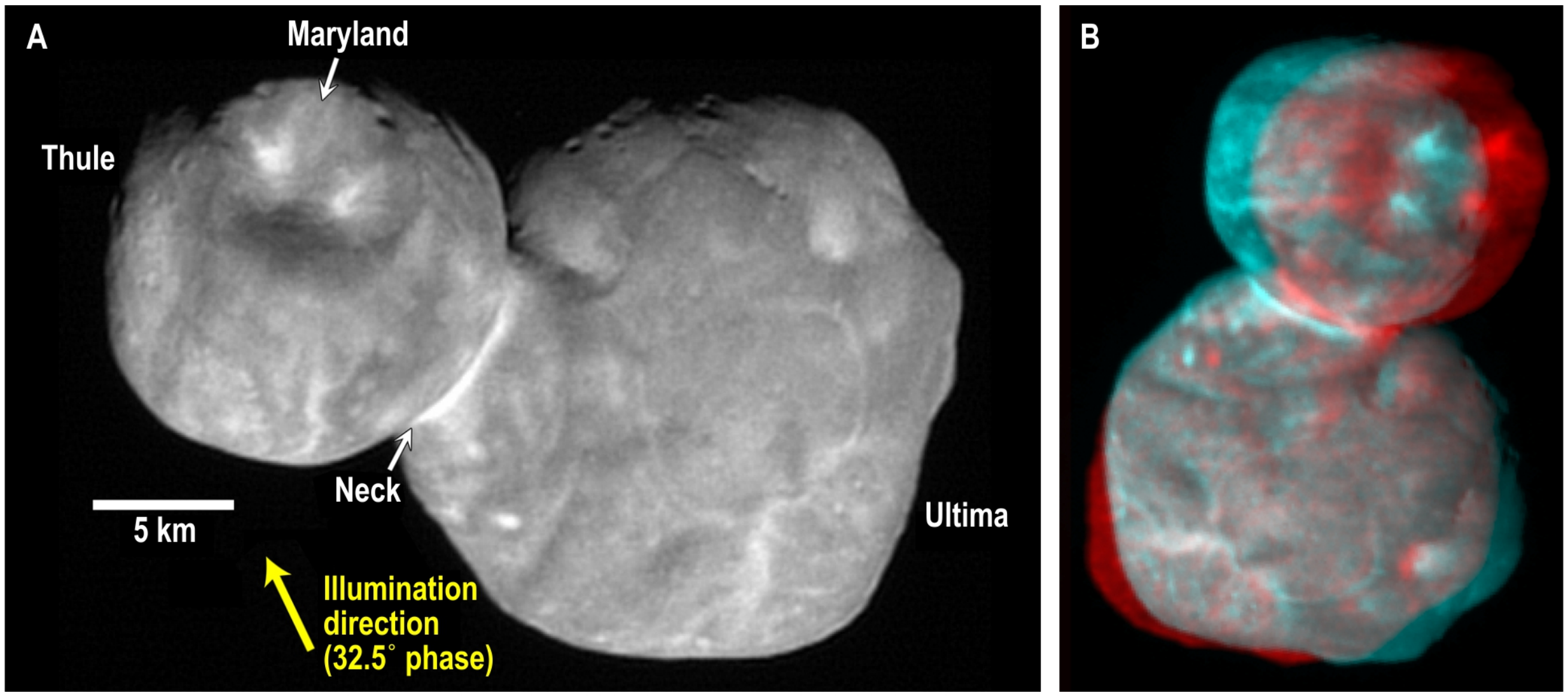

**Figure 1: . Close approach 06 (CA06) MVIC observation of $MU_{69}$.** Image A was made at a solar phase angle of 32.5°, a range of 6,640 km, and a native pixel scale of 130 m per pixel. A: This MVIC image was resampled to a 2x finer pixel scale, and then deconvolved using a similarly resampled MVIC point-spread function to create this figure. B: Red-cyan stereographic anaglyph of the $MU_{69}$ approach hemisphere.

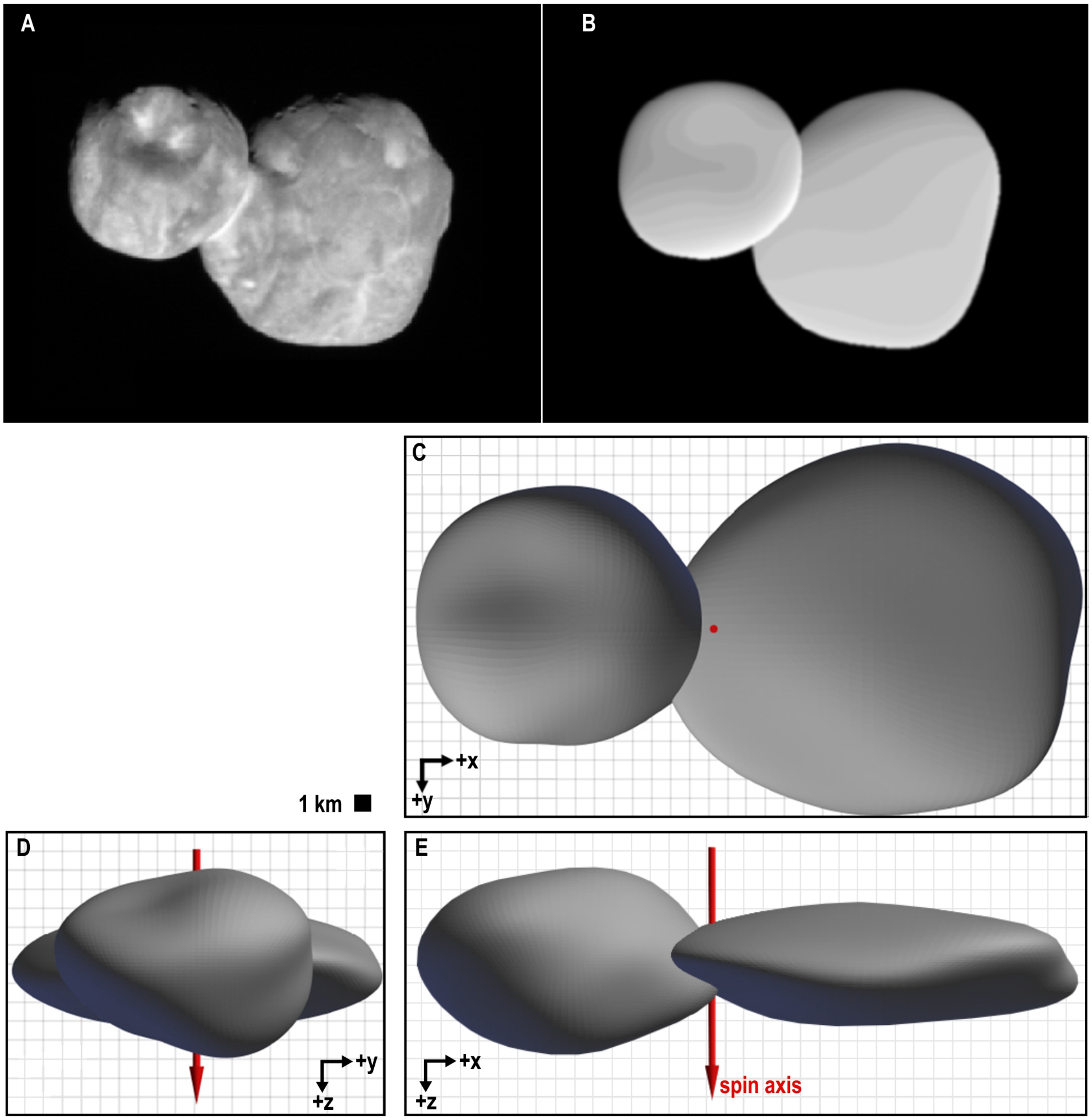

**Figure 2. Shape model for $MU_{69}$.** A: The same MVIC image as Fig 1B for comparison. B: Shape model of $MU_{69}$ based on approach rotational coverage and close flyby observations, seen at the same viewing and illumination conditions as A. C-E: Shape model seen along $MU_{69}$'s z (C), x (D), and y (E) axes, with the object's spin axis indicated in red. The arrow of the spin axis is the positive (i.e. north) pole. The shape of the +z hemisphere, which was mostly in darkness during the encounter, is the least well-constrained part of the shape model.

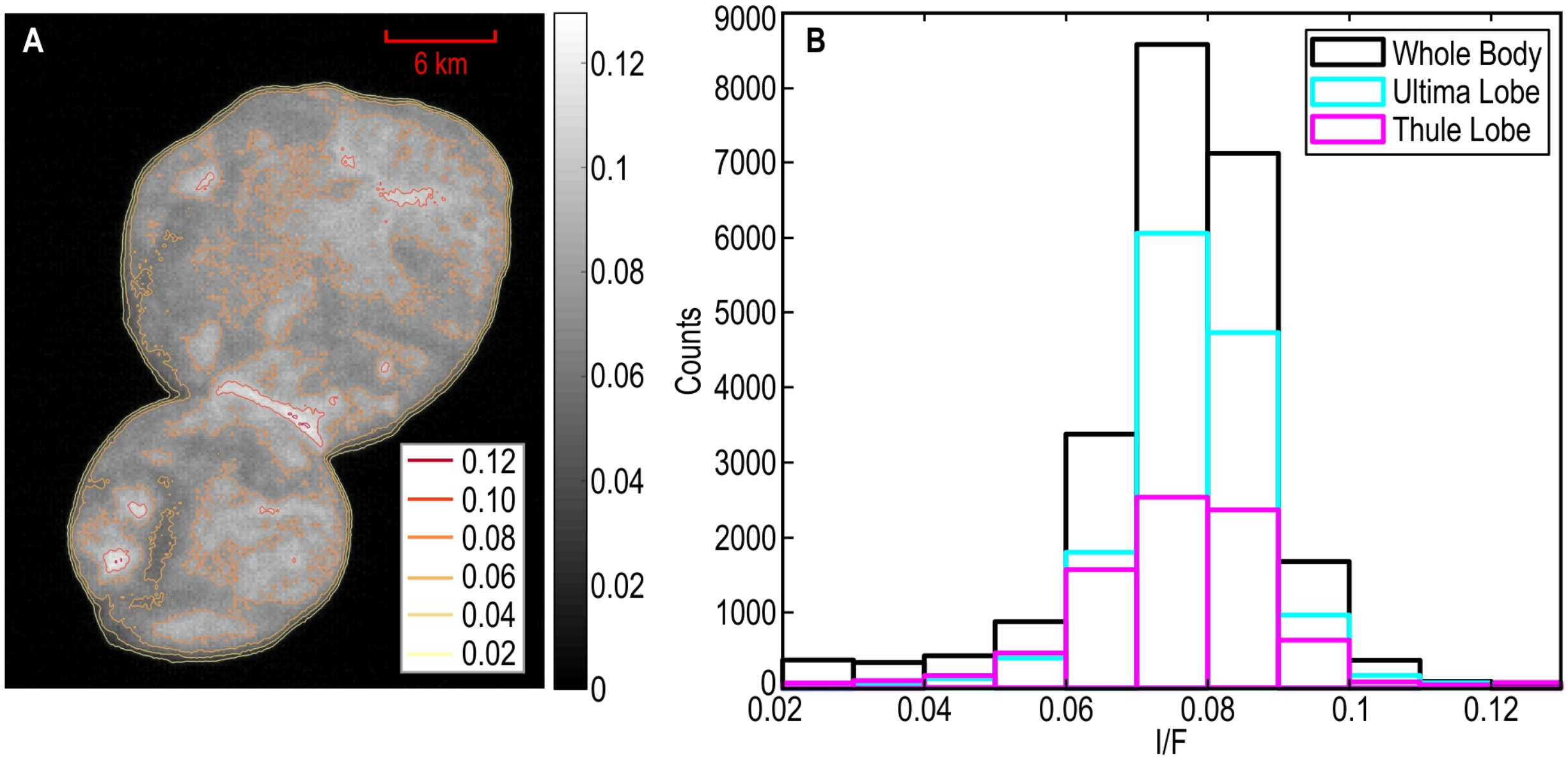

**Figure 3. Albedo contour map and histogram for the CA04 LORRI observation.** A: I/F isocontours of light scattered from $MU_{69}$'s surface at 13° solar phase angle. B: Histogram of I/F pixel values: These I/F values refer to the LORRI pivot wavelength (607.6 nm). The spectral distribution of $MU_{69}$ was approximated as a 50% Pluto plus 50% Pholus spectrum (see composition text section below).

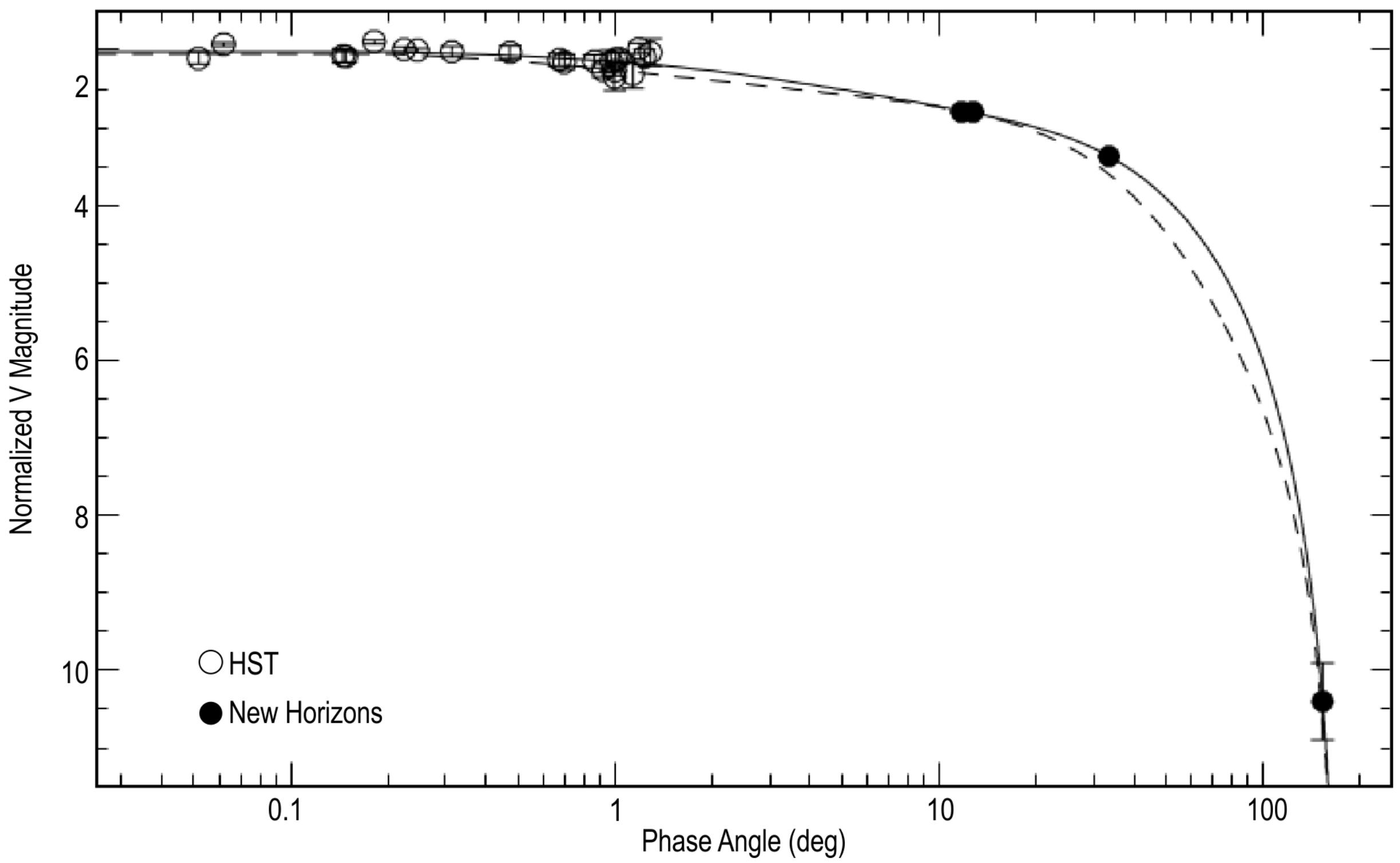

**Figure 4. The visible (0.55 μm) solar phase curve of $MU_{69}$ obtained by combining *HST* and *New Horizons* data.** The solid line is the Hapke photometric model (*81*) fitted to the total integrated I/F of $MU_{69}$. Data are from *New Horizons* close approach images acquired at solar phase angles of 12°, 13°, 32.5°, and 153°, and low phase angles (<1.3°) observations by the *Hubble Space Telescope*. Error bars on the 153° observation primarily arise from uncertainty in the shape of $MU_{69}$'s night side. Magnitudes are normalized to the geometric albedo ($p_V$=1) at opposition ($\alpha$=0 deg) in the *V* band (0.55 μm); no corrections have been made to account for rotational variation in reflectance (i.e., the lightcurve), but $MU_{69}$'s lightcurve amplitude upper limit is low (≤0.15 magnitudes [*10*]), so this neglected effect is small. For comparison, the dashed line is the solar phase curve of 103P/Hartley 2 (*82*), a bi-lobate Jupiter family comet visited by the *Deep Impact* spacecraft (*83*). Although Hartley 2 is darker ($p_V$=0.045±0.009) than $MU_{69}$ at 0.55 μm, the phase curves of these two bodies have similar shapes.

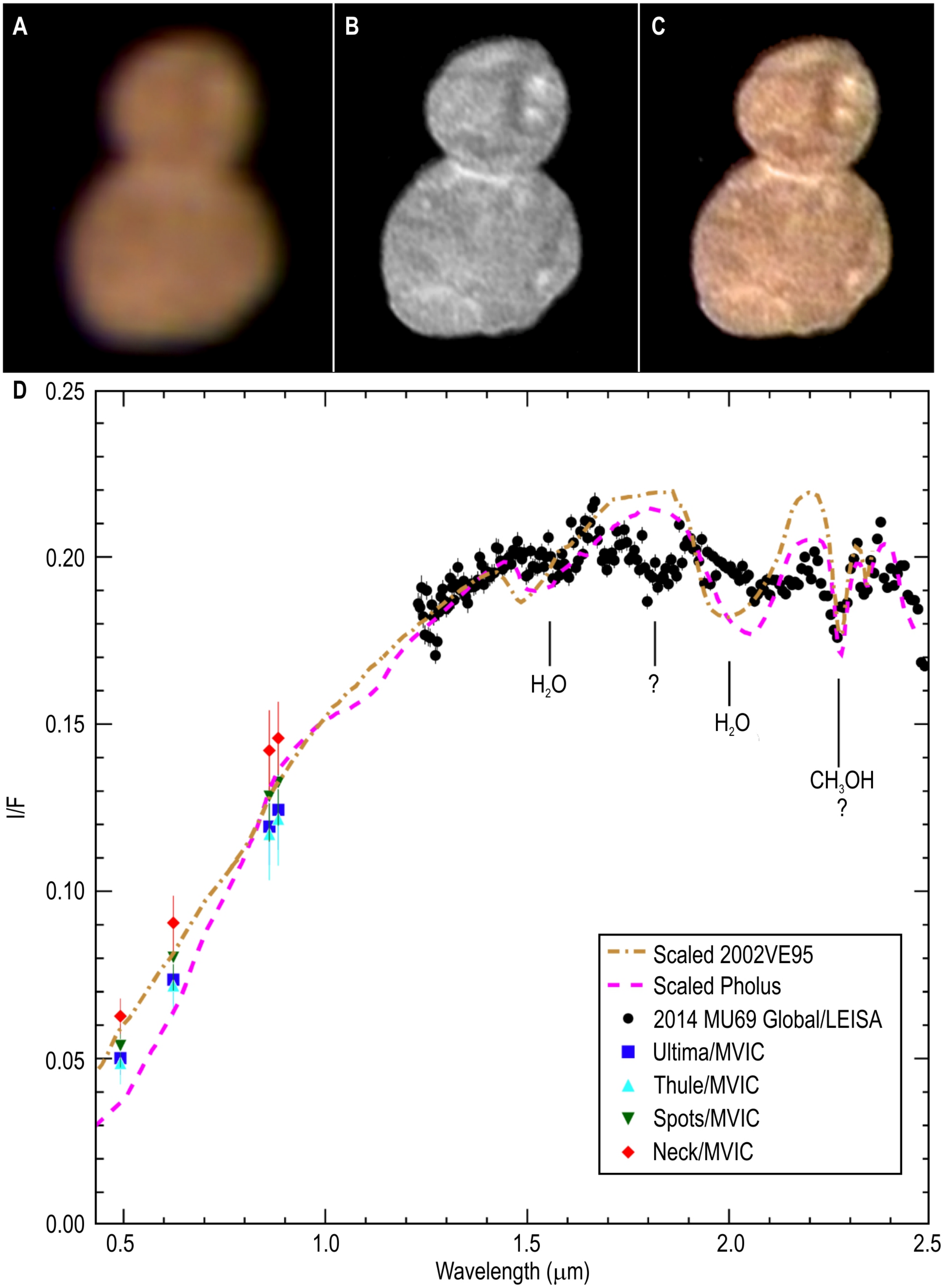

**Figure 5. $MU_{69}$'s color and near-IR spectral reflectance**. A: MVIC enhanced color image at 1.5 km per pixel scale, B: CA04 LORRI image at 140 m per pixel scale; C: A overlaid on B. D: MVIC color measurements (colored points) and a LEISA NIR spectrum of $MU_{69}$ (black points). Data at wavelengths shorter than 1 micron are from the MVIC visible/NIR color imager at a phase angle of 11.7°, while data longer than 1.2 μm are from the LEISA IR spectrograph at a phase angle of 12.6° and a mean spatial scale of 1.9 km per pixel. The MVIC data are split into multiple terrain units (Ultima and Thule lobes, the bright neck region, and a combination of all other bright spots identified in LORRI data), while the LEISA spectrum is a global average. All LEISA data points illustrate an estimated 1$\sigma$ uncertainty; MVIC data points illustrate an estimated 1$\sigma$ uncertainty relative to the red channel flux. The data are compared to Hapke model spectra shown as the brown dot-dashed line of 2002 $VE_{95}$ (*48*) and the magenta dashed line of 5145 Pholus (*49*). Those curves are scaled by 0.45 and 0.84, respectively, to match the average near IR I/F of $MU_{69}$. The apparent wavelength shifts of some features in the $MU_{69}$ spectrum relative to the dashed models are likely due to unmodeled temperature, particle size, and temperature effects. Tentative identifications of absorption bands of water and methanol ices are marked, along with an unknown feature at 1.8 μm (see text).

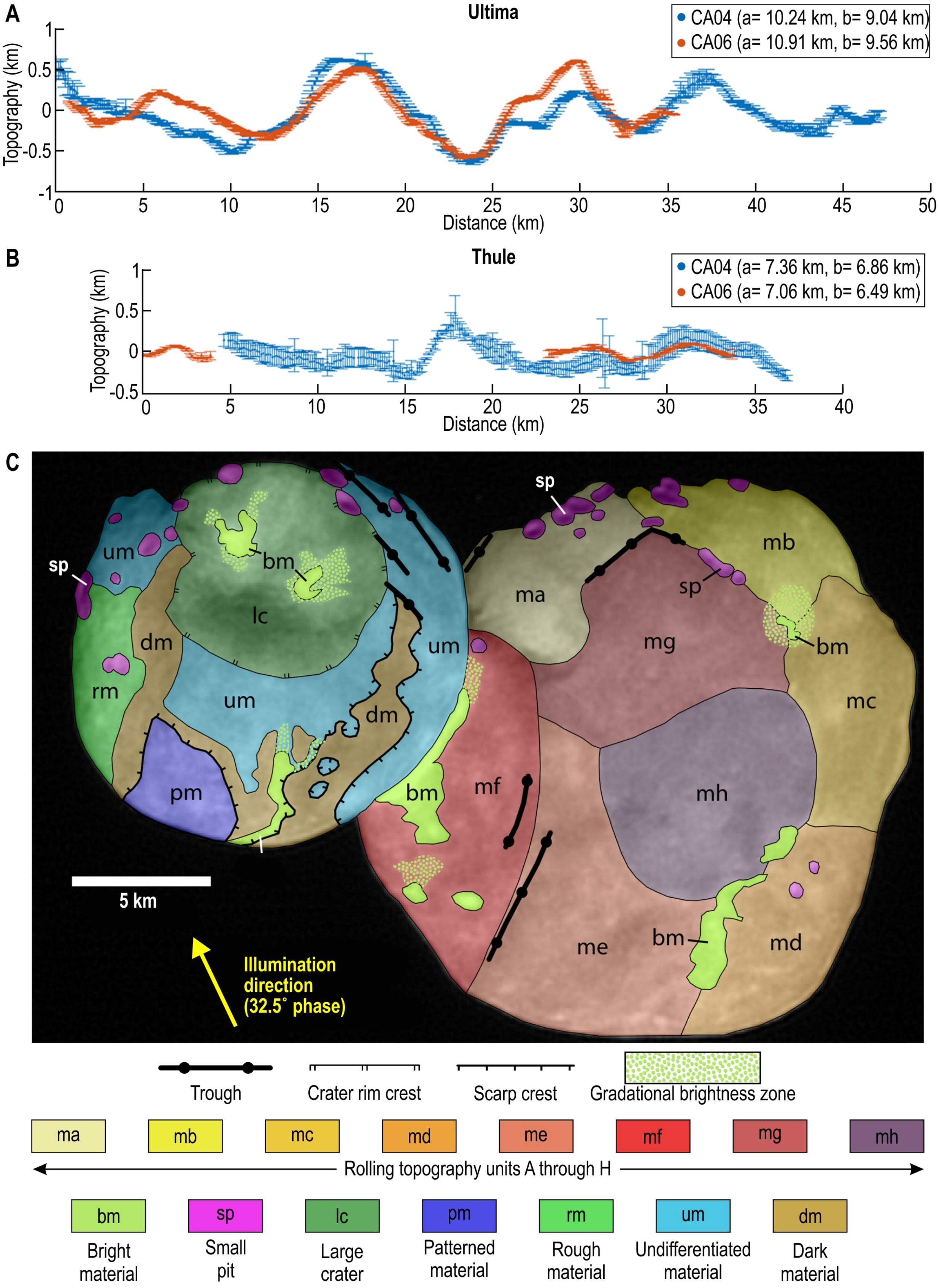

**Figure 6. Approach observation limb profiles and geomorphological map of $MU_{69}$.** A and B: Limb topography profiles of Ultima and Thule, respectively, measured using the LORRI CA04 and MVIC CA06 observations, after subtracting the best-fitting projected elliptical figures. The semi-major (a) and semi-minor (b) axes of the best-fitting ellipses are indicated in the legend. Error bars represent the difference in estimated limb positions between two independent workers; the median difference was ~0.3 pixels for $MU_{69}$ as a whole. The low solar phase angle of the CA04 observation (~12°) allows the limb to be measured more reliably around a larger portion of the perimeter of $MU_{69}$ than for CA06 (solar phase 32.5°). C: Geomorphological map of $MU_{69}$. The base map is the MVIC image from Fig. 2A. The positive spin axis of $MU_{69}$ is pointing approximately into the page. The mapped boundaries are preliminary. Note that this mapping is physiographic in nature and is not intended to rigorously convey stratigraphic relations between units.

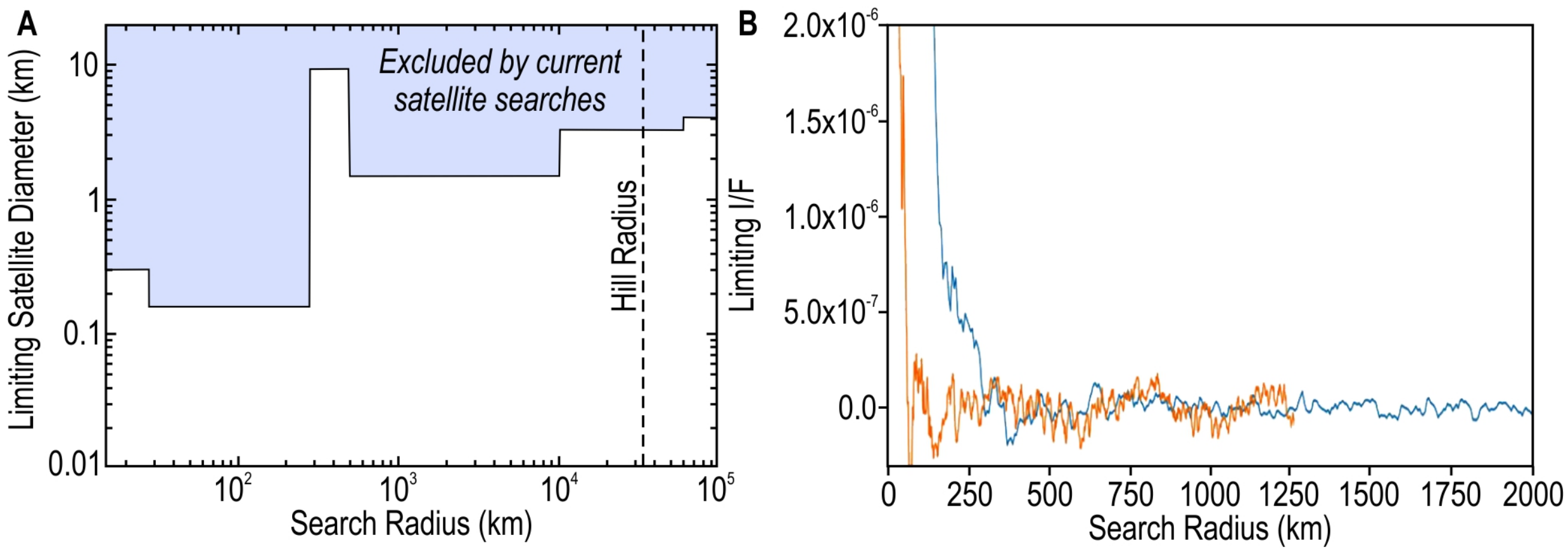

**Figure 7. Upper limits on satellites and ring/dust structures around $MU_{69}$.** A: Satellite search limiting diameters, assuming satellites have similar reflectivity to $MU_{69}$. The Hill radius (i.e., maximum orbit stability radius against solar tides) shown here assumes an effective spherical-equivalent radius of 7.5 km for $MU_{69}$ (approximating the lobes as ellipsoids) and a density of 500 kg $m^{-3}$. B: Profiles of limiting ring detectability for I/F versus distance from $MU_{69}$, assuming a 10 km wide ring, obtained using LORRI observations made at 22 (blue) and 6.5 (orange) hours before closest approach. The increasing I/F constraints inward of ~250 km are due to stray light from $MU_{69}$ within LORRI's optics, not material around $MU_{69}$.

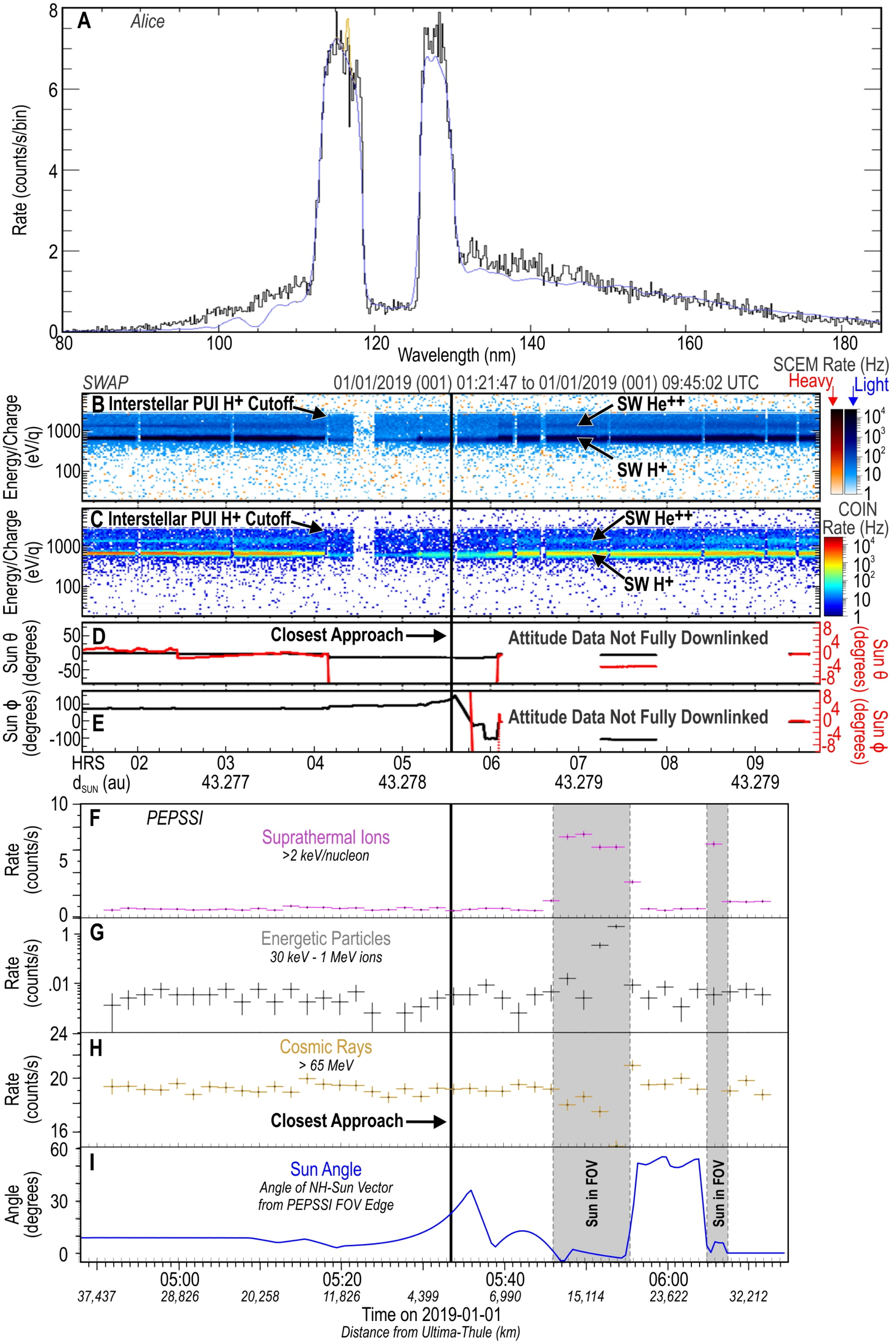

A
Alice
Rate (counts/s/bin)
Wavelength (nm)
SWAP
01/01/2019 (001) 01:21:47 to 01/01/2019 (001) 09:45:02 UTC
SCEM Rate (Hz)
Heavy
Light
B Interstellar PUI H+ Cutoff
SW He++
SW H+
Energy/Charge (eV/q)
C Interstellar PUI H+ Cutoff
COIN Rate (Hz)
D
Closest Approach
Attitude Data Not Fully Downlinked
Sun θ (degrees)
E
Sun φ (degrees)
HRS
dSUN (au)
F
PEPSSI
Suprathermal Ions
>2 keV/nucleon
Rate (counts/s)
G
Energetic Particles
30 keV - 1 MeV ions
H
Cosmic Rays
> 65 MeV
I
Sun Angle
Angle of NH-Sun Vector from PEPSSI FOV Edge
Angle (degrees)
Sun in FOV
Time on 2019-01-01
Distance from Ultima-Thule (km)

**Figure 8. $MU_{69}$ atmospheric and plasma search results**. A: Alice ultraviolet airglow spectrum. A modeled background (blue) includes interplanetary H emissions and four nearby stars. A small model signal at 116.6 nm (gold) indicates the H 121.6 nm count rate that would be expected if $MU_{69}$ were outgassing H atoms at $Q \sim 3\times10^{24}$ atoms $s^{-1}$; the 121.6 nm H signal appears at this wavelength because the observations were offset to avoid a low sensitivity region of the detector. No $MU_{69}$ coma emission is detected in the observed spectrum (black). B-C: SWAP low energy plasma spectrometer data showing light (blue) and heavy (orange) secondary channel electron multiplier (SCEM) count rate energy/charge spectra, where light (i.e., $H^+$ and $He^{++}$) and heavy ions are distinguished by their low and high secondary/primary electron ratios as they pass through SWAP's carbon foil (69), and coincidence rates (COIN) spectra. The heavy ions are sparse and not associated with close approach. D-E: SWAP orientation with respect to the Sun in altitude (θ) and azimuth (φ) angles (full range: black; zoomed: red). All the changes observed in the coincidence rate spectra in the second panel are associated with changes in spacecraft orientation; no changes related to the presence of $MU_{69}$ were observed. F-I: PEPSSI energetic particle spectrometer data. Count rates are shown for three products for ~80 minutes near closest approach to $MU_{69}$ (indicated by the vertical black line). Data were acquired in 1-s bins but averaged over 2-minute intervals to improve the signal to noise ratio. F: Suprathermal particles at >2 keV nucleon$^{-1}$ (dominated by interstellar pickup $He^+$); G: energetic protons (30 keV to 1 MeV); H: galactic cosmic rays (mostly >100 MeV protons); and I: the angle between PEPSSI aperture and the Sun. Changes in instrument orientation account for all the observed count rate variability; all particle rates shown are typical for the undisturbed interplanetary medium, with an increased photon background when the Sun is in the PEPSSI field of view (FOV).